\documentclass[a4paper,11pt]{article}
\usepackage{pos}
\usepackage{bigstrut}
\usepackage{hyperref} 
\usepackage{lineno}
\usepackage{subfigure}
\usepackage{wrapfig,enumitem}

\title{Combined fit above 0.1 EeV to the cosmic-ray spectrum and composition measured with the Pierre Auger Observatory}
 \ShortTitle{Combined fit to the cosmic-ray spectrum and composition}

\author*[ab]{Esteban Roulet}

\affiliation[a]{Centro At\'omico Bariloche, Comisi\'on Nacional de Energ\'\i a At\'omica}
\affiliation[b]{Consejo Nacional de Investigaciones Cient\'\i ficas y T\'ecnicas (CONICET)\\
Av. Bustillo 9500, R8402AGP, Bariloche, Argentina}

\onbehalf{for the Pierre Auger Collaboration$^c$}
\affiliation[c]{Observatorio Pierre Auger, Av.\ San Mart{\'\i}n Norte 304, 5613 Malarg\"ue, Argentina\\
Full author list: {\rm\url{https://www.auger.org/archive/authors_icrc_2025.html}}}

\emailAdd{spokespersons@auger.org}

\abstract{We fit the cosmic-ray spectrum measured with the Pierre Auger Observatory's surface detectors above an energy of $10^{17}$ eV, along with composition information inferred from the depth of shower maximum measured with its fluorescence detectors above an energy threshold of $10^{17.8}$ eV. We consider astrophysical scenarios with two distinct extragalactic source populations: one dominating the flux above a few EeV, and the other dominating at lower energies, with representative nuclei being injected at the sources with power-law spectra and rigidity-dependent cutoffs. The high-energy population exhibits a hard source injection spectrum and a relatively heavy composition, while the low-energy population exhibits a softer spectrum and a lighter composition. Extending the fit down to the low energies considered here shows the potential to test the energy region towards the Galactic to extragalactic transition with the data of the Pierre Auger Observatory. 
In particular, the Galactic contribution is expected to be still sizable at the lowest energies considered, and the extragalactic contribution needs to become suppressed for decreasing energies, an effect that could naturally result from a magnetic-horizon-induced suppression.}

\ConferenceLogo{PoS_ICRC2025_logo}

\FullConference{%
39th International Cosmic Ray Conference (ICRC2025)\\
15 -- 24 July, 2025\\
Geneva, Switzerland}

\begin{document}
\maketitle

\section{Introduction}

The cosmic-ray spectrum and the composition inferred from the measurements of the depth of shower maximum $X_{\rm max}$, have been jointly fitted by the Auger Collaboration using physically motivated astrophysical scenarios \cite{aa17,xcf,adcfit,cfwmh}. In these scenarios usually five representative mass components (H, He, N, Si and Fe) are considered to be emitted from uniformly distributed extragalactic sources,  having a common power-law spectra and rigidity-dependent cutoffs. The cosmological source evolution and the propagation effects through the radiation backgrounds are also accounted for in order to obtain the spectrum of the cosmic rays (CRs) arriving at Earth. 

The first studies \cite{aa17} considered energies above $10^{18.7}$\,eV and included just one population of extragalactic sources. The main conclusion was that this high-energy (HE) population was required to have hard source spectra (d$\Phi$/d$E\propto E^{-\gamma}$, with $\gamma_{\rm HE}<1$) and a relatively small cutoff rigidity, so that the CRs of charge $Z$ get suppressed for energies greater than a few $Z$\,EeV. In this way, the observed trend towards a heavier composition for increasing energies can be reproduced and each mass component becomes relevant only in a narrow energy range, in agreement with the small values of the dispersion $\sigma(X_{\rm max})$ measured above the ankle energy.

Extensions down to a threshold energy of $10^{17.8}$\,eV were then performed \cite{xcf}, which required the consideration of a second population with different composition and spectral properties and eventually also with different cosmological evolution. This low-energy (LE) component, which becomes dominant below a few EeV, contains mostly light and intermediate mass nuclei and is required to have a steep source spectrum ($\gamma_{\rm LE}>3.3$). It is presumably a new population of extragalactic origin, although the possibility that it could be partly related to a Galactic component and/or to secondary particles produced in the interactions of CR nuclei in the neighborhood of the sources was also discussed. 
Moreover, since this LE population extends above the ankle energy, the spectrum of the HE component is now required to be even harder to leave room for the LE component. 
It was also found that the shape of the cutoff adopted has direct implications on the spectral slope and the cutoff rigidity inferred for the HE population, with steeper cutoff shapes leading to softer source spectra and higher cutoff rigidities \cite{xcf,cfwmh}. 

In the present work we further extend the combined fit analysis to even lower energies, considering the spectral data down to $10^{17}$\,eV \cite{spectrum}, besides the measured $X_{\rm max}$ moments above $10^{17.8}$\,eV \cite{yu19}.\footnote{We do not include $X_{\rm max}$ data at lower energies given their still preliminary status.} We focus on scenarios with two different extragalactic populations, with the HE one dominating above few EeV and the LE one dominating at lower energies. Note that since the LE population has a steep spectrum, its extension to lower energies can lead to an overshooting of the spectrum below a few hundred PeV. To avoid this, we will hence consider the impact of a magnetic horizon effect (MHE) due to the presence of extragalactic magnetic fields \cite{le05,be07}. The MHE could affect the spectrum of the low-energy extragalactic component once the finite density of this source population is taken into account. This is due to the fact that for small enough energies, even for the closest sources the time required for the CR diffusive propagation up to Earth may exceed the source's lifetime.  This will then suppress the CR flux from extragalactic sources for decreasing energies, avoiding the overshooting of the spectrum. In addition, we also study the impact of the Galactic component, which is expected to be non-negligible in the energy range being now considered. We adopt for it the results of \cite{mo19a}, where also measurements from other experiments extending down to PeV energies could be reproduced using the same five mass groups and considering a rigidity-dependent break in their spectrum. In this scenario, the knee of the spectrum at around 3\,PeV  is associated with the H break and the break of Fe lies close to 100\,PeV. 

\section{The astrophysical scenario}

\subsection{The extragalactic source populations}
We consider two different source populations of extragalactic sources, injecting the five representative elements (H, He, N, Si and Fe) with a power-law spectrum having a rigidity dependent cutoff. The differential particle generation rate of each component of atomic number $Z$ and mass number $A$, per unit volume, energy and time, is then
\begin{equation}
    \tilde Q_{A}^a(E,z)=\tilde Q_0^a\xi^a(z) f^a_A\left(\frac{E}{E_0}\right)^{-\gamma_a}{\rm sech}\left(\left(\frac{E}{ZR^a_{\rm cut}}\right)^\Delta\right),
    \label{qa.eq}
\end{equation}
with $a = ({\rm LE, HE})$ identifying the population dominating at low and high energies respectively. For each population, the normalization $\tilde Q^a_0$ is the present total differential rate of CR emission per unit energy, volume and time, at the reference energy $E_0$ (smaller than the hydrogen cutoff  $R^a_{\rm cut}$), at which the relative source fractions of the different elements are $f^a_A$. 
The factor $\xi^a(z)$ parameterises the evolution of the emissivity as a function of the redshift $z$, for which we consider here the simplest case $\xi^a=1$ associated to steady non-evolving sources.
The parameter $\Delta$ characterizes the steepness of the cutoff suppression, and in this analysis we will consider the values $\Delta=1$ or 2. For $\Delta=1$ the shape is actually quite similar to a broken exponential cutoff, while for $\Delta=2$ it is steeper, and this case has also recently been found to fit well numerical results of CR acceleration in magnetic turbulence \cite{co24}.

To account for the interactions during propagation, we use the SimProp code \cite{al17} with the TALYS photodisintegration cross sections \cite{talys} and the Gilmore et al.~extragalactic background light model \cite{gi12}.

\subsection{The magnetic horizon of the LE component}
Given that extragalactic magnetic fields are ubiquitous in the Universe and that the extragalactic sources have a finite density, one expects that a magnetic horizon effect should suppress the fluxes of the extragalactic cosmic rays for decreasing energies. In particular, this effect could naturally be the cause for the required suppression of the extragalactic LE contribution below a few hundred PeV \cite{le05}. Considering a uniform turbulent extragalactic field with root mean square strength $B_{\rm rms}$ and coherence length $L_{\rm coh}$, one can introduce the critical rigidity for which the Larmor radius equals the coherence length as $R_{\rm crit}\equiv e B_{\rm rms}L_{\rm coh} \simeq 0.9(B_{\rm rms}/{\rm nG})(L_{\rm coh}/{\rm Mpc})$\,EeV. If the sources have an average density $n_{\rm s}$, characterized by the intersource distance $d_{\rm s}=n_{\rm s}^{-1/3}$, the flux suppression should depend on the parameters $R_{\rm crit}$ and on the normalized source distance $X_{\rm s}\equiv d_{\rm s}/\sqrt{L_{\rm coh}r_{\rm H}}\simeq (d_{\rm s}/10\,{\rm Mpc})\sqrt{25\,{\rm kpc}/L_{\rm coh}}$, with $r_{\rm H}=c/H_0$ being the Hubble radius. One typically finds that the MHE leads to a suppression of the spectrum by a factor 0.5 for rigidities $E/Z\sim X_{\rm s}R_{\rm crit}/4$ for non-evolving sources, and for a rigidity of about half that value for sources evolving like the star formation rate \cite{go21}.

In \cite{cfwmh} we considered a HE population having a low source density, smaller than $10^{-4}$\,Mpc$^{-3}$, for which the MHE suppression could occur around the ankle energy for He nuclei. This could help to explain the hardness of the observed spectrum of the high-energy population in a scenario in which the source spectral slope is more in line with the expectations from diffusive shock acceleration ($\gamma\simeq 2$).  Considering that the LE population may consist of a much larger number of less powerful sources,  their higher density would imply that their associated magnetic horizon would become relevant only at lower energies, closer to 0.1\,EeV for H and He nuclei. This effect can be quantified through a suppression function $G(E) \equiv J(E)/J(E)_{d_{\rm s} \rightarrow 0}$, given by the ratio between the actual flux at Earth from the discrete source distribution and the flux that would result in the limit of a continuous source distribution (with the same emissivity per unit volume).\footnote{For the case of a continuous source distribution, the magnetic fields  have no suppression effect due to the so-called {\em propagation theorem} \cite{al04}, and the interaction effects should be similar to those obtained in the case of rectilinear propagation.  }
 To implement this suppression, we adopt the parameterization obtained in \cite{go21} from the study of the results of simulations of CR propagation in turbulent magnetic fields. This suppression has the expression
\begin{equation}
    G(x) = \exp \left[- \left( \frac{a \,X_{\rm s}}{x + b\,(x/a)^\beta}\right)^{\alpha} \right],
    \label{gfactor}
\end{equation}
where  $x\equiv E/(ZR_{\rm crit})$.
The parameters $a,\, b,\, \alpha$ and $\beta$, tabulated in \cite{go21}, are sensitive to the assumed cosmological evolution of the source population emissivity as well as to whether the particles are primaries emitted at the sources or secondaries produced by photo-disintegration interactions during their propagation.
 Given that for the LE component the magnetic horizon effect appears typically at rigidities below few hundred PeV, the interaction effects of the extragalactic CRs during their propagation are small in this regime, and hence only the suppression of the primary elements turns out to be relevant. The parameters $R_{\rm crit}$ and $X_{\rm s}$ that specify the shape of the low-energy magnetic horizon-induced suppression will be determined from the fit. We considered values $X_{\rm s}\leq 2$, which typically correspond to source densities $n_{\rm s}>10^{-4}$\,Mpc$^{-3}(L_{\rm coh}/25\,{\rm kpc})^{3/2}$.

\subsection{The Galactic component}
The Galactic component is expected to dominate the cosmic-ray spectrum up to about 100~PeV, an energy close to that of the second-knee in the spectrum. 
In any case, the Galactic component should extend beyond this energy, progressively steepening but probably only becoming negligible at energies larger than 1\,EeV. The Galactic component is therefore expected to have an impact on the fit we perform above 0.1\,EeV. 

We have considered the Galactic component as derived in \cite{mo19a}, consisting of a superposition of the same five elements adopted for the other populations, which should be representative of the different cosmic-ray mass groups. Their spectra are taken as rigidity dependent smooth broken power-laws with a cutoff  at high rigidities, so that the total spectrum is  parameterized as
\begin{equation}
\frac{{\rm d}\Phi_{\rm G}}{{\rm d}E}=\sum_A\frac{{\rm d}\Phi^A_{\rm G}}{{\rm d}E}=\phi_{\rm G}\sum_A\ f_A \left(\frac{E}{\rm EeV}\right)^{-\gamma_1}\left[1+\left(\frac{E}{ZR^{\rm G}_{\rm b}}\right)^{\Delta\gamma/w}\right]^{-w}\,{\rm sech}\left(\frac{E}{ZR^{\rm G}_{\rm cut}}\right).
\label{galflux}
\end{equation}
Relying on direct measurements at energies around 10 to 100~TeV, the fractions of the different representative elements are taken as $f_{\rm H}=f_{\rm He}=0.35$, $f_{\rm N}=0.12$, $f_{\rm Si}=0.08$ and $f_{\rm Fe}=0.1$. A fit to CR data above 1~PeV \cite{mo19a} determined that each component has a break at an energy $ZR^{\rm G}_{\rm b}$, with $R^{\rm G}_{\rm b}\simeq 3.1$\,PeV. At this break energy the individual spectra change from a power index $\gamma_1\simeq 2.76$ to $\gamma_2=\gamma_1+\Delta\gamma\simeq 3.45$, and $w\simeq 0.11$ characterizes the width of this transition. The normalization of the Galactic spectrum is $\phi_{\rm G}=419 /({\rm km^2\, sr\, yr\, EeV})$.  These values are those obtained when including in the fit the Auger data and using the EPOS-LHC hadronic model, although the resulting Galactic spectral parameters turn out to be quite similar for the other hadronic models considered. Note that in \cite{mo19a} the Auger energy scale was rescaled by a factor of 1.07 to match the Telescope Array energy scale. Transforming back to the Auger energy should then require a rescaling of the energies of the different features by a factor $\xi=1/1.07$ (e.g. $R^{\rm G}_{\rm b}\to \xi R^{\rm G}_{\rm b}$), and also the flux normalization should be modified as $\phi_{\rm G}\to \xi^{\gamma_1-1}\phi_{\rm G}$. 
In \cite{mo19a} the cutoff rigidity $R^{\rm G}_{\rm cut}$ was found to be about 40\,PeV, but its value is expected to depend on the actual shape of the magnetic horizon suppression of the LE extragalactic population. We will hence determine it in the present analysis from the combined fit including all the different components. 

\section{Results}

 In order to reproduce the measurements by the Pierre Auger Observatory we fit the 17 parameters of the model (for each extragalactic population there are 4 independent elemental abundances, the spectral slope, cutoff rigidity and the flux normalization, together with the parameters $X_{\rm s}$ and $R_{\rm crit}$ determining the MHE, as well as the Galactic cutoff rigidity). For the spectrum we use the events detected by the Surface Detector array. For energies  above 2.5~EeV we considered the array with stations separated by 1500~m while for smaller energies the denser array with stations separated by 750~m is adopted \cite{spectrum}. 
 The mass composition is inferred by fitting the first two moments of the $X_{\max}$ distributions, measured using the Fluorescence Detector telescopes, above an energy of $10^{17.8}$\,eV
 \cite{yu19}. 
There are 32 energy bins for the spectrum and 19 energy bins for $X_{\rm max}$, which leads to a total of $N=70$ data points to fit by minimizing the $\chi^2$ function (with the number of degrees of freedom being 53).

\begin{table}[b]
\centering
{\small
\scalebox{0.85}{
\begin{tabular}[H]{ @{}c| c c c c c c c c| c c c c c  c c c @{}}
& \multicolumn{8}{c|}{EPOS-LHC} & \multicolumn{8}{c}{Sibyll\,2.3d}  \\
\hline
$\Delta$ &  $\gamma_{\rm HE}$ & $R_{\rm cut}^{\rm HE}$ & $\gamma_{\rm LE}$ & $R_{\rm cut}^{\rm LE}$ & $X_{\rm s}$ & $R_{\rm crit}$ & $R_{\rm cut}^{\rm G}$ & $\chi^2$ & $\gamma_{\rm HE}$ & $R_{\rm cut}^{\rm HE}$ & $\gamma_{\rm LE}$ & $R_{\rm cut}^{\rm LE}$  & $X_{\rm s}$ & $R_{\rm crit}$ & $R_{\rm cut}^{\rm G}$ & $\chi^2$\bigstrut[t]\\ 
&  &  [EeV] & &  [EeV] &  & [EeV]  & [PeV] &  & &  [EeV] & &  [EeV] &  & [EeV]   &[PeV] &  \\
\hline
1 &  -2.1 & 1.4 & 3.6 & $100$ & 2 & 0.19 & 25 &  98.4 &  -1.6 & 1.4 & 3.5 & 2.6 & 2 & 0.12& 15& 97.8\\
2 &  0.49 & 6.3 & 3.4 & 100 & 2 & 0.18 & 85 & 114 &  0.56 & 6.3 & 3.6 & 100 & 2 & 0.14 & 17 & 82.4\\
\end{tabular}}}

\caption{Parameters of the fit to the spectrum and $X_{\rm max}$ moments.  Results for the EPOS-LHC and Sibyll\,2.3d hadronic interaction models are reported, for cutoff shapes with $\Delta =1$ or 2. The corresponding  $\chi^2$ values are given. }
\label{table_1}
\end{table}

\begin{figure}[t]
    \centering
\includegraphics[width=0.4\textwidth]{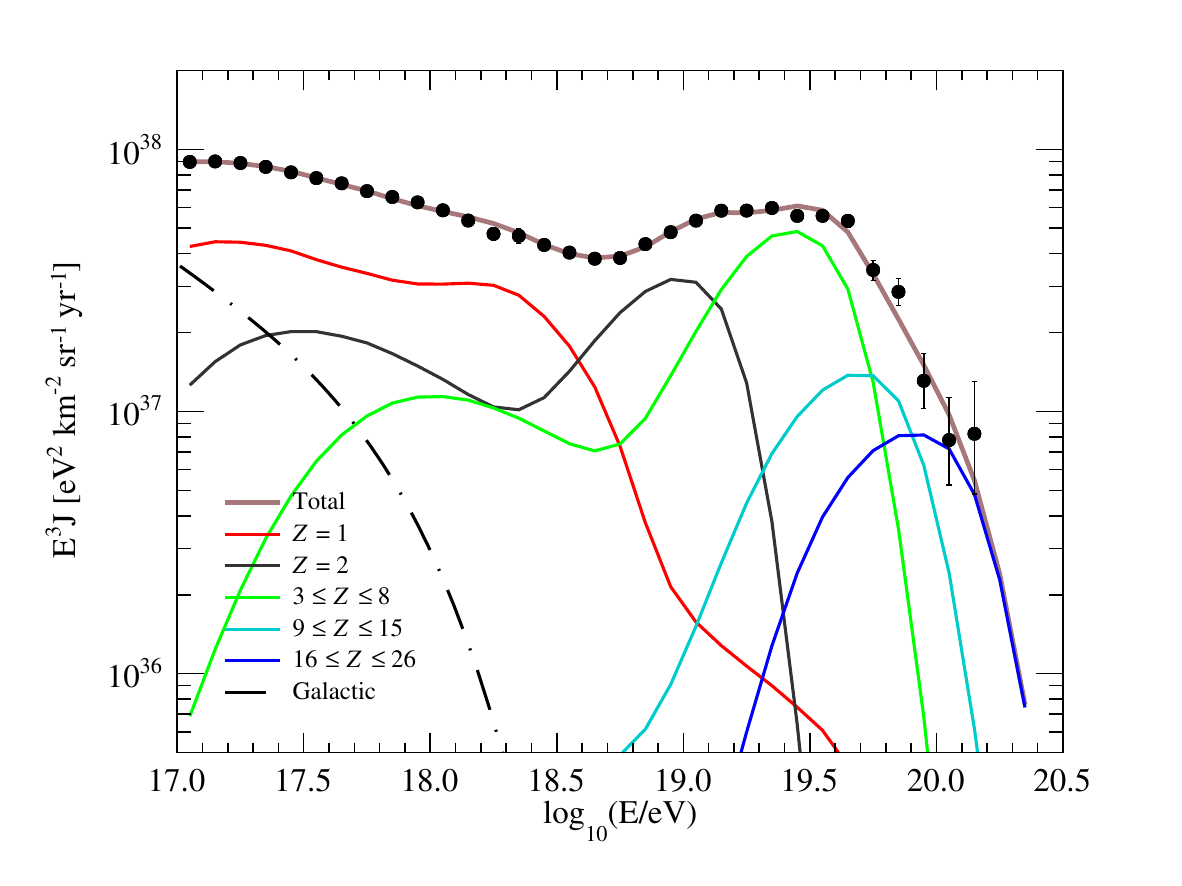}
    \includegraphics[width=0.4\textwidth]{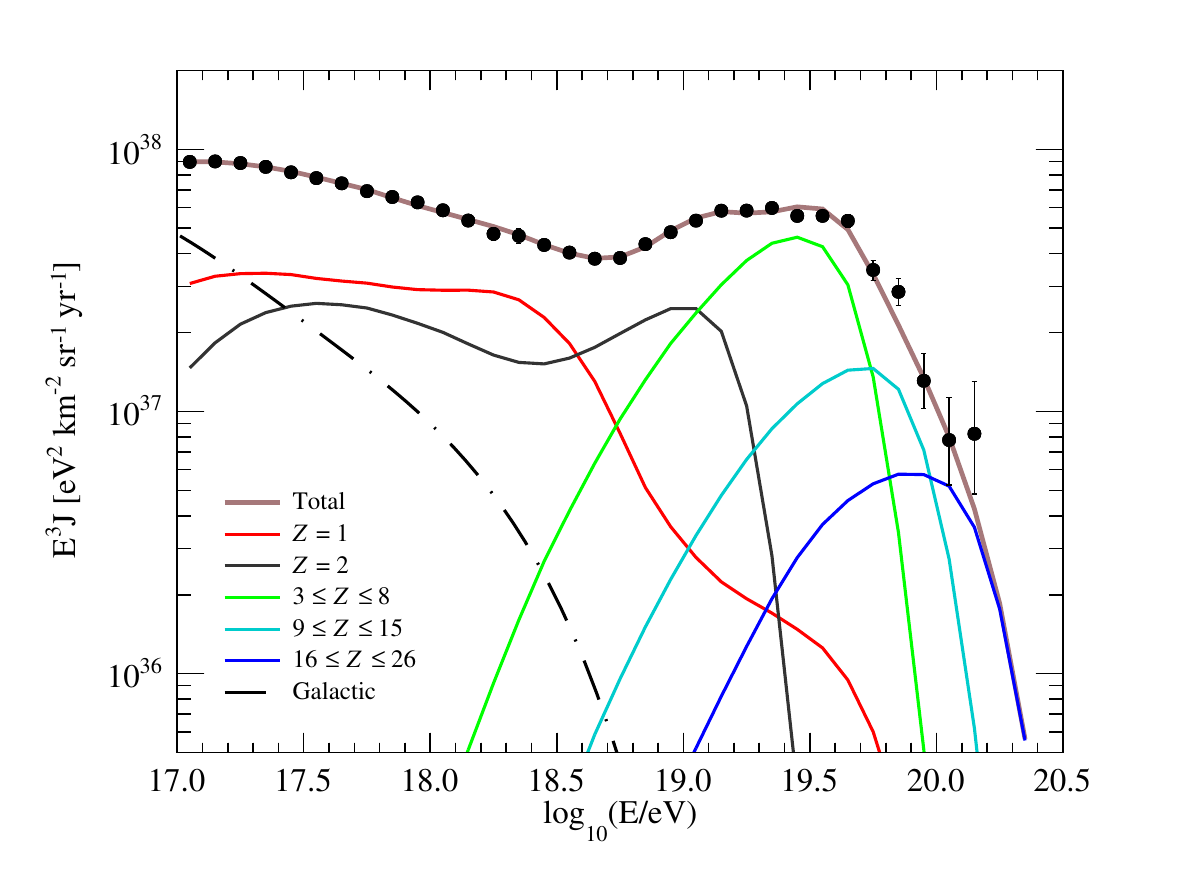}
    
   \subfigure[{\small $\Delta=1$, EPOS-LHC}]{ \includegraphics[width=0.485\textwidth]{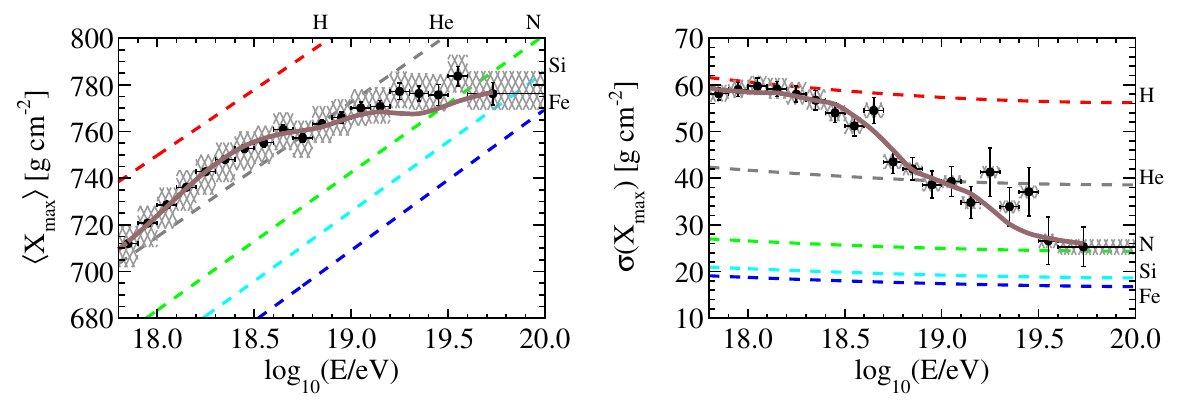}}
    \subfigure[{\small $\Delta=2$, EPOS-LHC}]{ \includegraphics[width=0.485\textwidth]{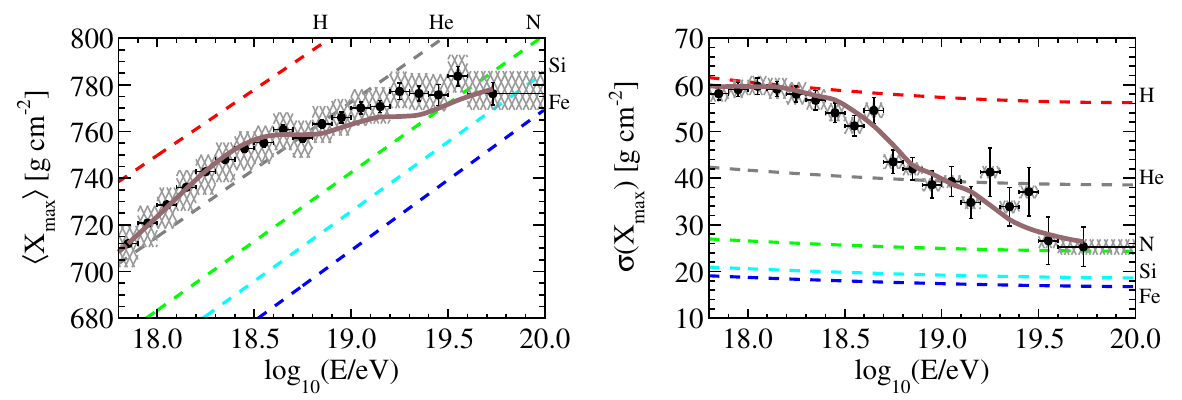}}
    \centering
\includegraphics[width=0.4\textwidth]{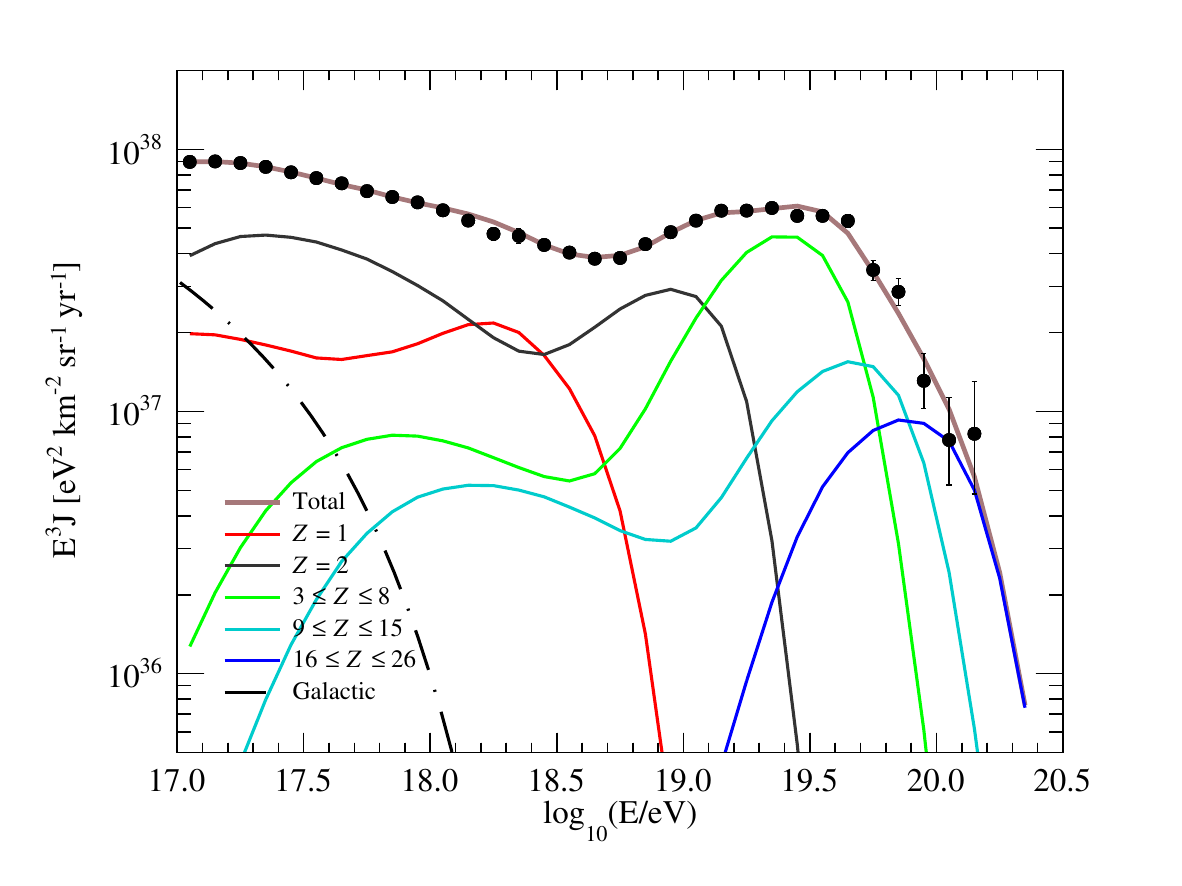}
    \includegraphics[width=0.4\textwidth]{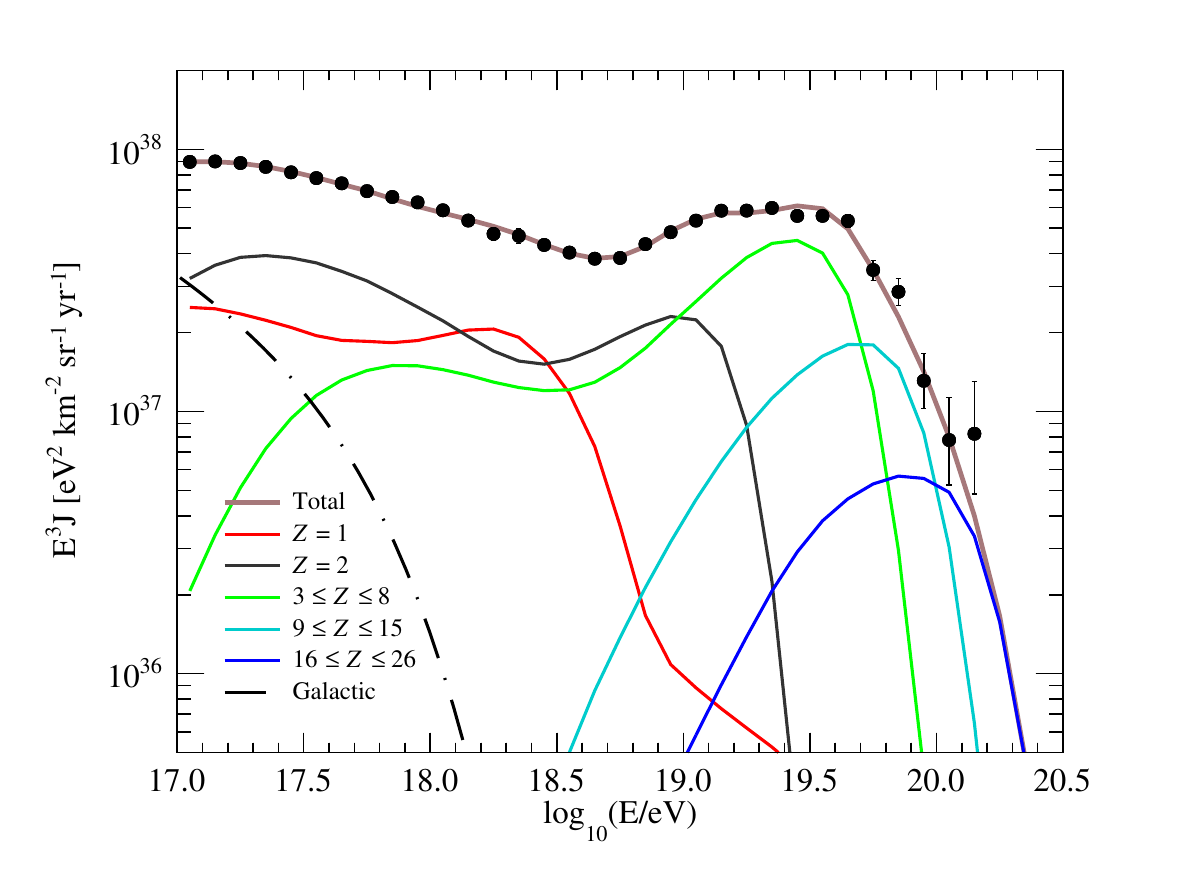}
    
   \subfigure[{\small $\Delta=1$, Sibyll\,2.3d}]{ \includegraphics[width=0.485\textwidth]{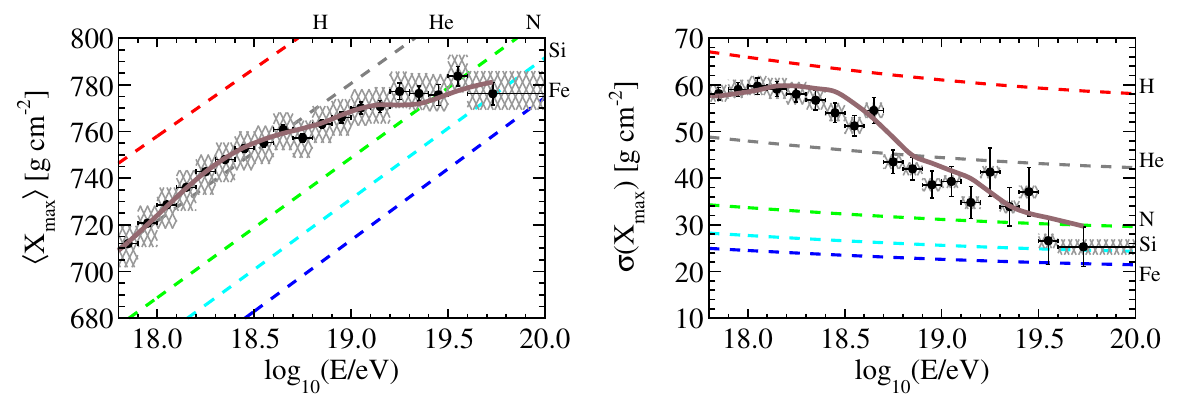}}
    \subfigure[{\small $\Delta=2$, Sibyll\,2.3d}]{ \includegraphics[width=0.485\textwidth]{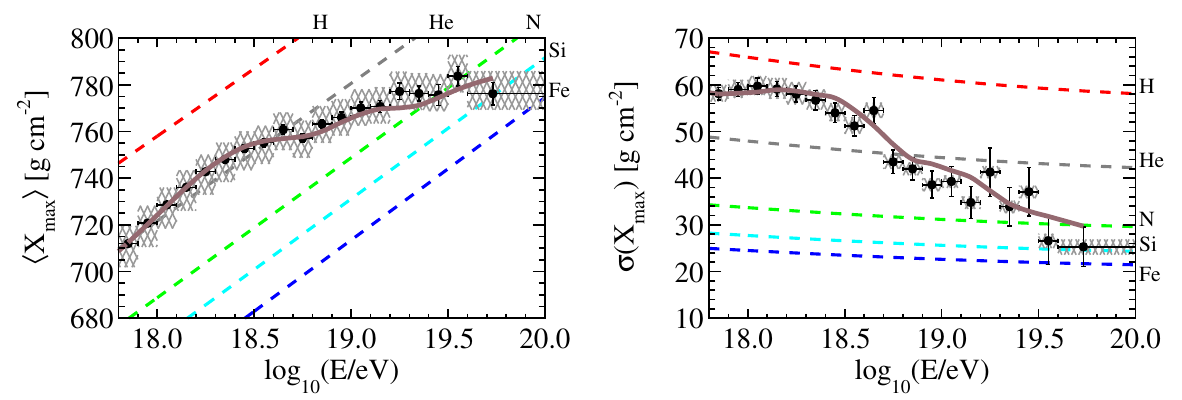}}
    \caption{Results of the fit to the spectrum for $E\geq10^{17}\,\rm{eV}$  and $X_{\rm max}$ moments for $E\geq 10^{17.8}\,\rm{eV}$. The top row is for the EPOS-LHC hadronic model while the bottom one for Sibyll\,2.3d. Left panels are for $\Delta=1$ while right panels for $\Delta=2$.  }
    \label{fig:spandmom}
\end{figure}

In Table~\ref{table_1} we show some of the results of the fit performed.  We considered the two hadronic interaction models,  EPOS-LHC and Sibyll\,2.3d, which are relevant for the determination of the cosmic-ray masses from the measured $X_{\rm max}$ moments.  We also adopt two different shapes for the cutoff function of the extragalactic component, considering $\Delta=1$ or 2. In Fig.~\ref{fig:spandmom} we display the measured spectrum and $X_{\rm max}$ moments, together with the results from the fitted models, which are found to reproduce the measurements reasonably well.

The general characteristics of the two extragalactic populations are similar to those found in previous works considering energies above $10^{17.8}$\,eV. The HE population has a hard spectrum with a rigidity cutoff of few EeV, leading to a succession of narrow bumps for each mass group and an increasingly heavy composition. The LE extragalactic population is instead lighter and has a steep source spectrum.
One can see that the composition turns out to be lighter with the EPOS-LHC hadronic model than with the Sibyll\,2.3d model. The shape of the cutoff has an impact on the reconstructed parameters of the HE component, with $\Delta=2$ leading to softer source spectra and higher rigidity cutoffs, although the overall shapes of the spectra at Earth are not very different in the two cases. For the Galactic CR model considered (from \cite{mo19a}), the cutoff rigidity is found to be between 15 and 85~PeV, depending on the particular hadronic interaction model adopted.

\begin{figure}[t]
    \centering
\includegraphics[width=0.45\textwidth]{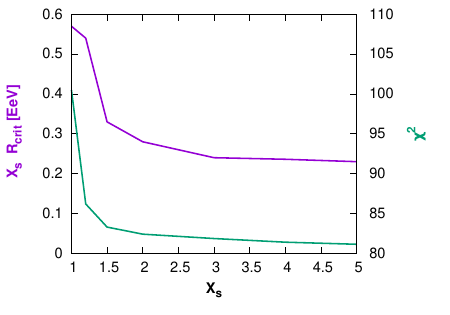}
    \includegraphics[width=0.4\textwidth]{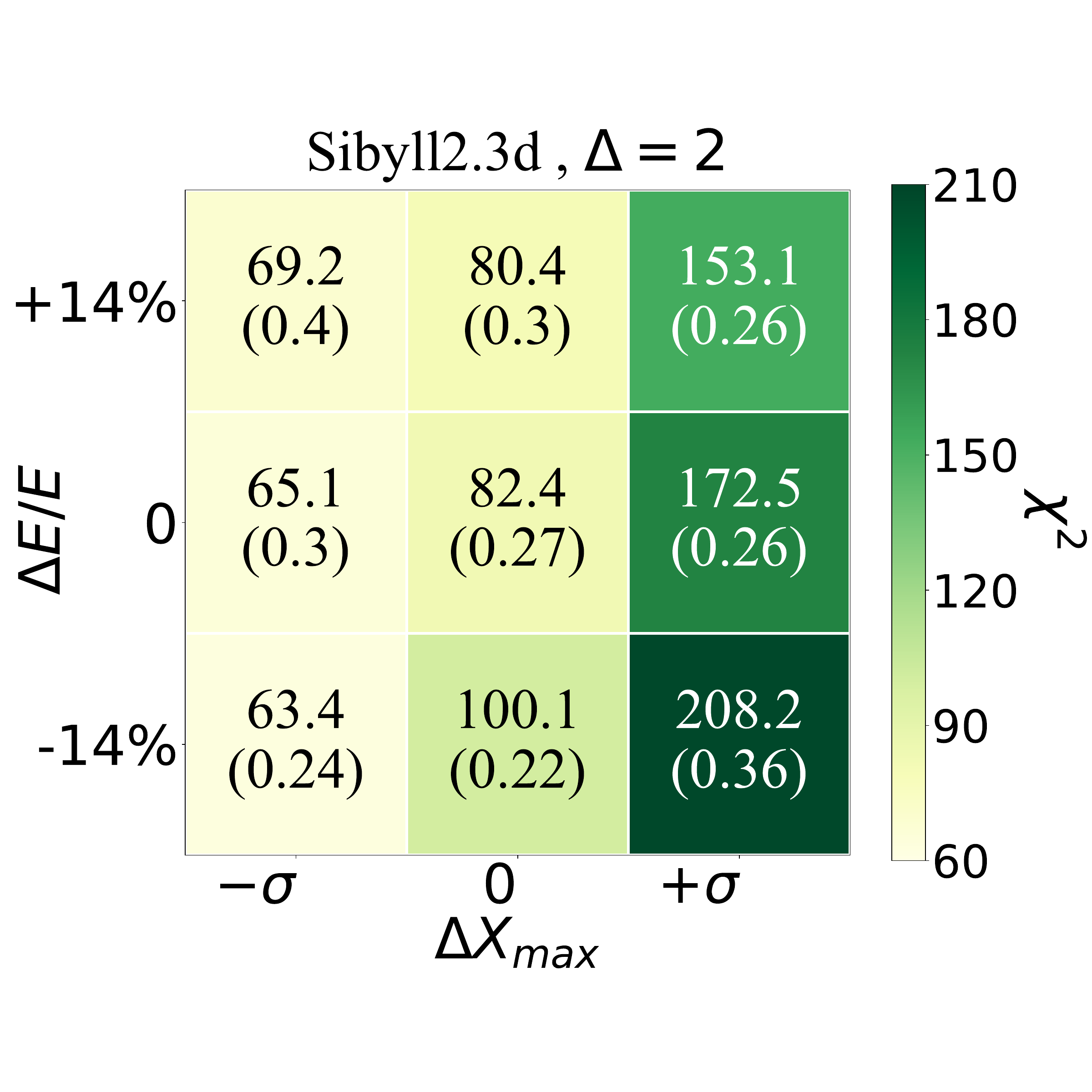}
    \caption{Left panel: Values of the $\chi^2$  (right scale) and of $X_{\rm s}R_{\rm crit}$ (left scale) as a function of $X_{\rm s}$. Right panel: effect of the systematic shifts on the values of the  $\chi^2$  and on $X_{\rm s}R_{\rm crit}$ (within parentheses). Both cases are for the Sibyll\,2.3d hadronic model with $\Delta=2$.  }
    \label{fig:2}
\end{figure}

In order not to overshoot the measured spectrum, the population dominating below the ankle needs to get suppressed below a few hundred PeV. This can happen naturally once the finite source density and the presence of extragalactic magnetic fields are taken into account, with the suppression resulting from the magnetic horizon effect.  In particular, the observed flattening of the spectrum occurring below 200\,PeV results in this scenario from the effect of the magnetic horizon on the light H and He extragalactic contributions from the LE population. This typically requires to have $X_{\rm s}R_{\rm crit}\simeq 0.3$\,EeV, which translates into a constraint in terms of the extragalactic magnetic field parameters and intersource distance
\begin{equation}
    B_{\rm rms}\simeq 13 \frac{10\,{\rm Mpc}}{d_{\rm s}}\sqrt{\frac{25\ {\rm kpc}}{L_{\rm coh}}}\,{\rm nG}.
\end{equation}
One should keep in mind that the relevant magnetic field strength is that in the region between Earth and the closest UHECR sources, i.e. within the Local Supercluster region, which is expected to be much larger than that in the voids of the large-scale structure.\footnote{The structure in the extragalactic magnetic field strength could introduce an additional ingredient in the analysis, so that the assumption of a uniform magnetic turbulence is an approximation to the average behavior expected in our local neighborhood.} 

Let us note that the normalized intersource distance $X_{\rm s}$ often slides towards the largest allowed value in the fit (equal to 2 in this analysis), but the $\chi^2$  remains very flat for larger values, as illustrated in the left panel of Fig.~\ref{fig:2}, where we also show that the product $X_{\rm s}R_{\rm crit}$ is almost constant for larger $X_{\rm s}$ values. Similarly, in many cases the cutoff of the LE population slides towards its highest allowed value of 100\,EeV, but the $\chi^2$ in these cases is actually very flat for $R_{\rm cut}^ {\rm LE}>30$\,EeV. 
Let us also mention that had we ignored the MHE, the fit would have been much worse, with the  $\chi^2$  typically increasing by more than 100 units.

On the other hand, systematic effects could possibly affect the energy calibration or the $X_{\rm max}$ values, and in the right panel of Fig. 2 we illustrate the changes in the  $\chi^2$  (and in the values of inferred for $X_{\rm s}R_{\rm crit}$ within parentheses) after performing positive or negative shifts by one systematic deviation on those quantities.  Note that the systematic uncertainty in the energy is about 14\%, while that in $X_{\rm max}$ is energy dependent, being about 10\,g\,cm$^{-2}$ at 1~EeV and about 7\,g\,cm$^{-2}$ at 10~EeV. One finds that negative shifts in $X_{\rm max}$, which lead to heavier inferred compositions,  usually improve the quality of the fit, while the value of $X_{\rm s}R_{\rm crit}$ is not strongly affected.

The presence of the tail of the heavier Galactic component helps in general to reproduce the observations, with the transition from a Galactic dominance to an extragalactic dominance taking place in this scenario at energies of about 0.1\,EeV. 
Extending the composition measurements to these energies, as well as extending the spectrum measurements using the denser array of surface detectors with separation of 433\,m, should help to better determine the features of the Galactic to extragalactic transition that were observed in the present work.

\clearpage

\section*{The Pierre Auger Collaboration}
\small

\begin{wrapfigure}[8]{l}{0.11\linewidth}
\vspace{-5mm}
\includegraphics[width=0.98\linewidth]{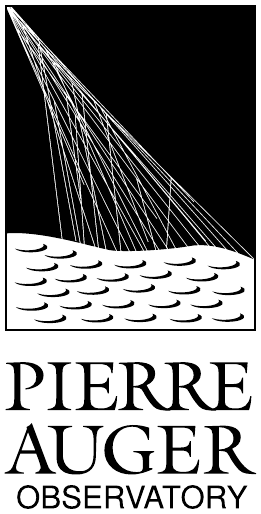}
\end{wrapfigure}
\begin{sloppypar}\noindent
A.~Abdul Halim$^{13}$,
P.~Abreu$^{70}$,
M.~Aglietta$^{53,51}$,
I.~Allekotte$^{1}$,
K.~Almeida Cheminant$^{78,77}$,
A.~Almela$^{7,12}$,
R.~Aloisio$^{44,45}$,
J.~Alvarez-Mu\~niz$^{76}$,
A.~Ambrosone$^{44}$,
J.~Ammerman Yebra$^{76}$
% ...

\end{sloppypar}

\begin{center}
\rule{0.1\columnwidth}{0.5pt}
\raisebox{-0.4ex}{\scriptsize$\bullet$}
\rule{0.1\columnwidth}{0.5pt}
\end{center}

\vspace{-1ex}
\footnotesize
\begin{description}[labelsep=0.2em,align=right,labelwidth=0.7em,labelindent=0em,leftmargin=2em,noitemsep,before={\renewcommand\makelabel[1]{##1 }}]
\item[$^{1}$] Centro At\'omico Bariloche and Instituto Balseiro (CNEA-UNCuyo-CONICET), San Carlos de Bariloche, Argentina
\item[$^{2}$] Departamento de F\'\i{}sica and Departamento de Ciencias de la Atm\'osfera y los Oc\'eanos, FCEyN, Universidad de Buenos Aires and CONICET, Buenos Aires, Argentina
\item[$^{3}$] IFLP, Universidad Nacional de La Plata and CONICET, La Plata, Argentina
\item[$^{4}$] Instituto de Astronom\'\i{}a y F\'\i{}sica del Espacio (IAFE, CONICET-UBA), Buenos Aires, Argentina
\item[$^{5}$] Instituto de F\'\i{}sica de Rosario (IFIR) -- CONICET/U.N.R.\ and Facultad de Ciencias Bioqu\'\i{}micas y Farmac\'euticas U.N.R., Rosario, Argentina
\item[$^{6}$] Instituto de Tecnolog\'\i{}as en Detecci\'on y Astropart\'\i{}culas (CNEA, CONICET, UNSAM), and Universidad Tecnol\'ogica Nacional -- Facultad Regional Mendoza (CONICET/CNEA), Mendoza, Argentina
\item[$^{7}$] Instituto de Tecnolog\'\i{}as en Detecci\'on y Astropart\'\i{}culas (CNEA, CONICET, UNSAM), Buenos Aires, Argentina
\item[$^{8}$] International Center of Advanced Studies and Instituto de Ciencias F\'\i{}sicas, ECyT-UNSAM and CONICET, Campus Miguelete -- San Mart\'\i{}n, Buenos Aires, Argentina
\item[$^{9}$] Laboratorio Atm\'osfera -- Departamento de Investigaciones en L\'aseres y sus Aplicaciones -- UNIDEF (CITEDEF-CONICET), Argentina
\item[$^{10}$] Observatorio Pierre Auger, Malarg\"ue, Argentina
\item[$^{11}$] Observatorio Pierre Auger and Comisi\'on Nacional de Energ\'\i{}a At\'omica, Malarg\"ue, Argentina
\item[$^{12}$] Universidad Tecnol\'ogica Nacional -- Facultad Regional Buenos Aires, Buenos Aires, Argentina
\item[$^{13}$] University of Adelaide, Adelaide, S.A., Australia
\item[$^{14}$] Universit\'e Libre de Bruxelles (ULB), Brussels, Belgium
\item[$^{15}$] Vrije Universiteit Brussels, Brussels, Belgium
\item[$^{16}$] Centro Brasileiro de Pesquisas Fisicas, Rio de Janeiro, RJ, Brazil
\item[$^{17}$] Centro Federal de Educa\c{c}\~ao Tecnol\'ogica Celso Suckow da Fonseca, Petropolis, Brazil
\item[$^{18}$] Instituto Federal de Educa\c{c}\~ao, Ci\^encia e Tecnologia do Rio de Janeiro (IFRJ), Brazil
\item[$^{19}$] Universidade de S\~ao Paulo, Escola de Engenharia de Lorena, Lorena, SP, Brazil
\item[$^{20}$] Universidade de S\~ao Paulo, Instituto de F\'\i{}sica de S\~ao Carlos, S\~ao Carlos, SP, Brazil
\item[$^{21}$] Universidade de S\~ao Paulo, Instituto de F\'\i{}sica, S\~ao Paulo, SP, Brazil
\item[$^{22}$] Universidade Estadual de Campinas (UNICAMP), IFGW, Campinas, SP, Brazil
\item[$^{23}$] Universidade Estadual de Feira de Santana, Feira de Santana, Brazil
\item[$^{24}$] Universidade Federal de Campina Grande, Centro de Ciencias e Tecnologia, Campina Grande, Brazil
\item[$^{25}$] Universidade Federal do ABC, Santo Andr\'e, SP, Brazil
\item[$^{26}$] Universidade Federal do Paran\'a, Setor Palotina, Palotina, Brazil
\item[$^{27}$] Universidade Federal do Rio de Janeiro, Instituto de F\'\i{}sica, Rio de Janeiro, RJ, Brazil
\item[$^{28}$] Universidad de Medell\'\i{}n, Medell\'\i{}n, Colombia
\item[$^{29}$] Universidad Industrial de Santander, Bucaramanga, Colombia
\item[$^{30}$] Charles University, Faculty of Mathematics and Physics, Institute of Particle and Nuclear Physics, Prague, Czech Republic
\item[$^{31}$] Institute of Physics of the Czech Academy of Sciences, Prague, Czech Republic
\item[$^{32}$] Palacky University, Olomouc, Czech Republic
\item[$^{33}$] CNRS/IN2P3, IJCLab, Universit\'e Paris-Saclay, Orsay, France
\item[$^{34}$] Laboratoire de Physique Nucl\'eaire et de Hautes Energies (LPNHE), Sorbonne Universit\'e, Universit\'e de Paris, CNRS-IN2P3, Paris, France
\item[$^{35}$] Univ.\ Grenoble Alpes, CNRS, Grenoble Institute of Engineering Univ.\ Grenoble Alpes, LPSC-IN2P3, 38000 Grenoble, France
\item[$^{36}$] Universit\'e Paris-Saclay, CNRS/IN2P3, IJCLab, Orsay, France
\item[$^{37}$] Bergische Universit\"at Wuppertal, Department of Physics, Wuppertal, Germany
\item[$^{38}$] Karlsruhe Institute of Technology (KIT), Institute for Experimental Particle Physics, Karlsruhe, Germany
\item[$^{39}$] Karlsruhe Institute of Technology (KIT), Institut f\"ur Prozessdatenverarbeitung und Elektronik, Karlsruhe, Germany
\item[$^{40}$] Karlsruhe Institute of Technology (KIT), Institute for Astroparticle Physics, Karlsruhe, Germany
\item[$^{41}$] RWTH Aachen University, III.\ Physikalisches Institut A, Aachen, Germany
\item[$^{42}$] Universit\"at Hamburg, II.\ Institut f\"ur Theoretische Physik, Hamburg, Germany
\item[$^{43}$] Universit\"at Siegen, Department Physik -- Experimentelle Teilchenphysik, Siegen, Germany
\item[$^{44}$] Gran Sasso Science Institute, L'Aquila, Italy
\item[$^{45}$] INFN Laboratori Nazionali del Gran Sasso, Assergi (L'Aquila), Italy
\item[$^{46}$] INFN, Sezione di Catania, Catania, Italy
\item[$^{47}$] INFN, Sezione di Lecce, Lecce, Italy
\item[$^{48}$] INFN, Sezione di Milano, Milano, Italy
\item[$^{49}$] INFN, Sezione di Napoli, Napoli, Italy
\item[$^{50}$] INFN, Sezione di Roma ``Tor Vergata'', Roma, Italy
\item[$^{51}$] INFN, Sezione di Torino, Torino, Italy
\item[$^{52}$] Istituto di Astrofisica Spaziale e Fisica Cosmica di Palermo (INAF), Palermo, Italy
\item[$^{53}$] Osservatorio Astrofisico di Torino (INAF), Torino, Italy
\item[$^{54}$] Politecnico di Milano, Dipartimento di Scienze e Tecnologie Aerospaziali , Milano, Italy
\item[$^{55}$] Universit\`a del Salento, Dipartimento di Matematica e Fisica ``E.\ De Giorgi'', Lecce, Italy
\item[$^{56}$] Universit\`a dell'Aquila, Dipartimento di Scienze Fisiche e Chimiche, L'Aquila, Italy
\item[$^{57}$] Universit\`a di Catania, Dipartimento di Fisica e Astronomia ``Ettore Majorana``, Catania, Italy
\item[$^{58}$] Universit\`a di Milano, Dipartimento di Fisica, Milano, Italy
\item[$^{59}$] Universit\`a di Napoli ``Federico II'', Dipartimento di Fisica ``Ettore Pancini'', Napoli, Italy
\item[$^{60}$] Universit\`a di Palermo, Dipartimento di Fisica e Chimica ''E.\ Segr\`e'', Palermo, Italy
\item[$^{61}$] Universit\`a di Roma ``Tor Vergata'', Dipartimento di Fisica, Roma, Italy
\item[$^{62}$] Universit\`a Torino, Dipartimento di Fisica, Torino, Italy
\item[$^{63}$] Benem\'erita Universidad Aut\'onoma de Puebla, Puebla, M\'exico
\item[$^{64}$] Unidad Profesional Interdisciplinaria en Ingenier\'\i{}a y Tecnolog\'\i{}as Avanzadas del Instituto Polit\'ecnico Nacional (UPIITA-IPN), M\'exico, D.F., M\'exico
\item[$^{65}$] Universidad Aut\'onoma de Chiapas, Tuxtla Guti\'errez, Chiapas, M\'exico
\item[$^{66}$] Universidad Michoacana de San Nicol\'as de Hidalgo, Morelia, Michoac\'an, M\'exico
\item[$^{67}$] Universidad Nacional Aut\'onoma de M\'exico, M\'exico, D.F., M\'exico
\item[$^{68}$] Institute of Nuclear Physics PAN, Krakow, Poland
\item[$^{69}$] University of \L{}\'od\'z, Faculty of High-Energy Astrophysics,\L{}\'od\'z, Poland
\item[$^{70}$] Laborat\'orio de Instrumenta\c{c}\~ao e F\'\i{}sica Experimental de Part\'\i{}culas -- LIP and Instituto Superior T\'ecnico -- IST, Universidade de Lisboa -- UL, Lisboa, Portugal
\item[$^{71}$] ``Horia Hulubei'' National Institute for Physics and Nuclear Engineering, Bucharest-Magurele, Romania
\item[$^{72}$] Institute of Space Science, Bucharest-Magurele, Romania
\item[$^{73}$] Center for Astrophysics and Cosmology (CAC), University of Nova Gorica, Nova Gorica, Slovenia
\item[$^{74}$] Experimental Particle Physics Department, J.\ Stefan Institute, Ljubljana, Slovenia
\item[$^{75}$] Universidad de Granada and C.A.F.P.E., Granada, Spain
\item[$^{76}$] Instituto Galego de F\'\i{}sica de Altas Enerx\'\i{}as (IGFAE), Universidade de Santiago de Compostela, Santiago de Compostela, Spain
\item[$^{77}$] IMAPP, Radboud University Nijmegen, Nijmegen, The Netherlands
\item[$^{78}$] Nationaal Instituut voor Kernfysica en Hoge Energie Fysica (NIKHEF), Science Park, Amsterdam, The Netherlands
\item[$^{79}$] Stichting Astronomisch Onderzoek in Nederland (ASTRON), Dwingeloo, The Netherlands
\item[$^{80}$] Universiteit van Amsterdam, Faculty of Science, Amsterdam, The Netherlands
\item[$^{81}$] Case Western Reserve University, Cleveland, OH, USA
\item[$^{82}$] Colorado School of Mines, Golden, CO, USA
\item[$^{83}$] Department of Physics and Astronomy, Lehman College, City University of New York, Bronx, NY, USA
\item[$^{84}$] Michigan Technological University, Houghton, MI, USA
\item[$^{85}$] New York University, New York, NY, USA
\item[$^{86}$] University of Chicago, Enrico Fermi Institute, Chicago, IL, USA
\item[$^{87}$] University of Delaware, Department of Physics and Astronomy, Bartol Research Institute, Newark, DE, USA
\item[] -----
\item[$^{a}$] Max-Planck-Institut f\"ur Radioastronomie, Bonn, Germany
\item[$^{b}$] also at Kapteyn Institute, University of Groningen, Groningen, The Netherlands
\item[$^{c}$] School of Physics and Astronomy, University of Leeds, Leeds, United Kingdom
\item[$^{d}$] Fermi National Accelerator Laboratory, Fermilab, Batavia, IL, USA
\item[$^{e}$] Pennsylvania State University, University Park, PA, USA
\item[$^{f}$] Colorado State University, Fort Collins, CO, USA
\item[$^{g}$] Louisiana State University, Baton Rouge, LA, USA
\item[$^{h}$] now at Graduate School of Science, Osaka Metropolitan University, Osaka, Japan
\item[$^{i}$] Institut universitaire de France (IUF), France
\item[$^{j}$] now at Technische Universit\"at Dortmund and Ruhr-Universit\"at Bochum, Dortmund and Bochum, Germany
\end{description}

\vspace{-1ex}
\footnotesize
\section*{Acknowledgments}

\begin{sloppypar}
The successful installation, commissioning, and operation of the Pierre
Auger Observatory would not have been possible without the strong
commitment and effort from the technical and administrative staff in
Malarg\"ue. We are very grateful to the following agencies and
organizations for financial support:
\end{sloppypar}

\begin{sloppypar}
Argentina -- Comisi\'on Nacional de Energ\'\i{}a At\'omica; Agencia Nacional de
Promoci\'on Cient\'\i{}fica y Tecnol\'ogica (ANPCyT); Consejo Nacional de
Investigaciones Cient\'\i{}ficas y T\'ecnicas (CONICET); Gobierno de la
Provincia de Mendoza; Municipalidad de Malarg\"ue; NDM Holdings and Valle
Las Le\~nas; in gratitude for their continuing cooperation over land
access; Australia -- the Australian Research Council; Belgium -- Fonds
de la Recherche Scientifique (FNRS); Research Foundation Flanders (FWO),
Marie Curie Action of the European Union Grant No.~101107047; Brazil --
Conselho Nacional de Desenvolvimento Cient\'\i{}fico e Tecnol\'ogico (CNPq);
Financiadora de Estudos e Projetos (FINEP); Funda\c{c}\~ao de Amparo \`a
Pesquisa do Estado de Rio de Janeiro (FAPERJ); S\~ao Paulo Research
Foundation (FAPESP) Grants No.~2019/10151-2, No.~2010/07359-6 and
No.~1999/05404-3; Minist\'erio da Ci\^encia, Tecnologia, Inova\c{c}\~oes e
Comunica\c{c}\~oes (MCTIC); Czech Republic -- GACR 24-13049S, CAS LQ100102401,
MEYS LM2023032, CZ.02.1.01/0.0/0.0/16{\textunderscore}013/0001402,
CZ.02.1.01/0.0/0.0/18{\textunderscore}046/0016010 and
CZ.02.1.01/0.0/0.0/17{\textunderscore}049/0008422 and CZ.02.01.01/00/22{\textunderscore}008/0004632;
France -- Centre de Calcul IN2P3/CNRS; Centre National de la Recherche
Scientifique (CNRS); Conseil R\'egional Ile-de-France; D\'epartement
Physique Nucl\'eaire et Corpusculaire (PNC-IN2P3/CNRS); D\'epartement
Sciences de l'Univers (SDU-INSU/CNRS); Institut Lagrange de Paris (ILP)
Grant No.~LABEX ANR-10-LABX-63 within the Investissements d'Avenir
Programme Grant No.~ANR-11-IDEX-0004-02; Germany -- Bundesministerium
f\"ur Bildung und Forschung (BMBF); Deutsche Forschungsgemeinschaft (DFG);
Finanzministerium Baden-W\"urttemberg; Helmholtz Alliance for
Astroparticle Physics (HAP); Helmholtz-Gemeinschaft Deutscher
Forschungszentren (HGF); Ministerium f\"ur Kultur und Wissenschaft des
Landes Nordrhein-Westfalen; Ministerium f\"ur Wissenschaft, Forschung und
Kunst des Landes Baden-W\"urttemberg; Italy -- Istituto Nazionale di
Fisica Nucleare (INFN); Istituto Nazionale di Astrofisica (INAF);
Ministero dell'Universit\`a e della Ricerca (MUR); CETEMPS Center of
Excellence; Ministero degli Affari Esteri (MAE), ICSC Centro Nazionale
di Ricerca in High Performance Computing, Big Data and Quantum
Computing, funded by European Union NextGenerationEU, reference code
CN{\textunderscore}00000013; M\'exico -- Consejo Nacional de Ciencia y Tecnolog\'\i{}a
(CONACYT) No.~167733; Universidad Nacional Aut\'onoma de M\'exico (UNAM);
PAPIIT DGAPA-UNAM; The Netherlands -- Ministry of Education, Culture and
Science; Netherlands Organisation for Scientific Research (NWO); Dutch
national e-infrastructure with the support of SURF Cooperative; Poland
-- Ministry of Education and Science, grants No.~DIR/WK/2018/11 and
2022/WK/12; National Science Centre, grants No.~2016/22/M/ST9/00198,
2016/23/B/ST9/01635, 2020/39/B/ST9/01398, and 2022/45/B/ST9/02163;
Portugal -- Portuguese national funds and FEDER funds within Programa
Operacional Factores de Competitividade through Funda\c{c}\~ao para a Ci\^encia
e a Tecnologia (COMPETE); Romania -- Ministry of Research, Innovation
and Digitization, CNCS-UEFISCDI, contract no.~30N/2023 under Romanian
National Core Program LAPLAS VII, grant no.~PN 23 21 01 02 and project
number PN-III-P1-1.1-TE-2021-0924/TE57/2022, within PNCDI III; Slovenia
-- Slovenian Research Agency, grants P1-0031, P1-0385, I0-0033, N1-0111;
Spain -- Ministerio de Ciencia e Innovaci\'on/Agencia Estatal de
Investigaci\'on (PID2019-105544GB-I00, PID2022-140510NB-I00 and
RYC2019-027017-I), Xunta de Galicia (CIGUS Network of Research Centers,
Consolidaci\'on 2021 GRC GI-2033, ED431C-2021/22 and ED431F-2022/15),
Junta de Andaluc\'\i{}a (SOMM17/6104/UGR and P18-FR-4314), and the European
Union (Marie Sklodowska-Curie 101065027 and ERDF); USA -- Department of
Energy, Contracts No.~DE-AC02-07CH11359, No.~DE-FR02-04ER41300,
No.~DE-FG02-99ER41107 and No.~DE-SC0011689; National Science Foundation,
Grant No.~0450696, and NSF-2013199; The Grainger Foundation; Marie
Curie-IRSES/EPLANET; European Particle Physics Latin American Network;
and UNESCO.
\end{sloppypar}

\end{document}